\documentclass[fleqn,usenatbib,useAMS,twocolumn,a4paper, aps]{revtex4}

\usepackage{aas_macros}
\usepackage{mathtools}
\usepackage{amsmath, amssymb}
\usepackage[colorlinks = true, linkcolor=blue, citecolor=blue, urlcolor=cyan]{hyperref}
\usepackage[T1]{fontenc}
\usepackage{ae,aecompl}
\usepackage{hyperref}
\usepackage{soul}
\usepackage{xcolor}
\usepackage{ragged2e}

\usepackage{graphicx}	
\usepackage{amsmath}	
\usepackage{amssymb}	
\usepackage{bm}		
\usepackage{pdflscape}	

\begin{document}

\title [Massloss from magnetized disk]{Study of mass outflow rates from magnetized advective accretion disk around rotating black holes}

\author{Camelia Jana}
\email{camelia\_jana@iitg.ac.in}

\author{Santabrata Das}
\email{sbdas@iitg.ac.in}

\affiliation{Department of Physics, Indian Institute of Technology Guwahati, Guwahati, 781039, Assam, India.}

\begin{abstract}
	We develop and discuss a model formalism to study the properties of mass outflows that are emerged out from a relativistic, magnetized, viscous, advective accretion flow around a rotating black hole. In doing so, we consider the toroidal component as the dominant magnetic fields and synchrotron process is the dominant cooling mechanism inside the accretion disk. With this, we self-consistently solve the coupled accretion-ejection governing equations in the steady state and obtain the shock-induced global inflow-outflow solutions in terms of the inflow parameters, namely plasma-$\beta$ ($=p_{\rm gas}/p_{\rm mag}$, $p_{\rm gas}$ and $p_{\rm mag}$ being gas and magnetic pressures), accretion rates ($\dot m$) and viscosity ($\alpha_{\rm B}$), respectively. Using these solutions, we compute the mass outflow rate ($R_{\dot m}$, the ratio of outflow to inflow mass flux) and find that mass loss from the magnetized accretion disk continues to take place for wide range of inflow parameters and black hole spin ($a_{\rm k}$). We also observe that $R_{\dot m}$ strongly depends on plasma-$\beta$, $\dot m$, $\alpha_{\rm B}$ and $a_{\rm k}$, and it increases as the magnetic activity inside the accretion disk is increased. Further, we compute the maximum mass outflow rate ($R^{\rm max}_{\dot m}$) by freely varying the inflow parameters and find that for magnetic pressure dominated disk, $R^{\rm max}_{\dot m} \sim 24\%$ ($\sim 30\%$) for $a_{\rm k}=0.0$ ($0.99$). Finally, while discussing the implication of our model formalism, we compute the maximum jet kinetic power using $R^{\rm max}_{\dot m}$ which appears to be in close agreement with the observed jet kinetic power of several black hole sources.
\end{abstract}
	
	\maketitle
	

\section{Introduction}

The signature of jets/outflows are often observed in accreting black hole systems of all mass scales starting from X-ray binaries \cite[]{Mirabel-Rodriguez1994,Mirabel-Rodrguez1999,Hjellming-Rupen1995,Miller-Jones-etal2023} to active galactic nuclei \cite[AGNs;][]{Jennison-DasGupta1953,Junor-etal1999,Doeleman-etal2012,Blandford-etal2019}. Indeed, jets/outflows are expected to originate from the accreting matter itself as the black holes do not emit matter or radiation. Furthermore, \cite{Junor-etal1999} indicated that jets are emerged out from the vicinity of the central source of M87 and these findings are further supplemented by \cite{Blandford-etal2019}. 

Meanwhile, observational studies ascertain that the launching of jets/outflows is possibly linked with the spectral states of the accreting matter around black hole X-ray binaries \cite[BH-XRBs;][]{Vadawale-etal2001,Gallo-etal2003,Fender-etal-2009,Rushton-etal2010,Miller-etal2012,Radhika-etal-2016,Blandford-etal2019}. In particular, steady and powerful jets/outflows are commonly observed in the low-hard states (LHS) and hard-intermediate states (HIMS). On contrary, transient relativistic jets are generally seen in the soft-intermediate states (SIMS) \cite[]{Fender-etal-2004,Fender-etal-2009}. In the high-soft states (HSS), jets are not observed \cite[]{Fender-1999,Fender-etal-2004,Radhika-etal-2016}. All these findings evidently indicate that the ejection of matter is correlated with the presence of Compton corona (hereafter post-shock corona, PSC) and hence, it is highly likely that jets/outflows are launched from the inner part of the disc. This conjecture seems reasonable as \cite{Radhika-Nandi-2014,Radhika-etal-2016} pointed out the disk-jet coupling in explaining the spectro-temporal properties of the outbursting BH-XRBs.

Interestingly, it is indeed apparent that the jets/outflows are originated from the disk itself, however the exact physical mechanisms responsible for jet generation still remain elusive. Intuitively, it is reasonable to consider that the extreme gravity of the central source plays an important role in launching as well powering the jets. The seminal work of \cite{Blandford-Znajek-1977} (BZ) demonstrated the electromagnetic energy extraction mechanism involving magnetic fields around rotating black holes and indicated that such mechanism is viable to power the jets. In addition, \cite{Blanford-Payne-1982} (BP) also showed that energy and angular momentum of the infalling matter is magnetically removed by the field line and eventually carried off by the outgoing matter. Further, extensive numerical simulations of magnetohydrodynamic (MHD) accretion flow in both non-relativistic and relativistic regimes also confirm that jets/outflows are produced from accretion disk \cite[]{Shibata-etal1986,Koide-etal1999,DeVillers-etal2005,Hawley-etal2006, Ohsuga-Mineshige2011, Tchekhovskoy-etal-2011,Dihingia-etal2021, Kwan-etal2023,Aktar-etal2024}. In particular, \cite{Shibata-etal1986} reported that magnetic fields help in jet formation and its collimation process. \cite{Koide-etal1999} studied the magnetically driven relativistic jet from Schwardschild black hole which is found to be of two-layered shell structure. \cite{DeVillers-etal2005} examined the unbound outflows from accretion disk in Kerr space-time and found that inflowing matter is largely expelled by the centrifugal barrier, whereas black hole rotation does not influence the matter ejection although spin enhances the outflow strength \cite[]{Hawley-etal2006}. In case of radiatively inefficient flow, outflows are also seen to emerge out due to the combined effects of magnetic as well as gas pressures \cite[]{Ohsuga-Mineshige2011}. Further, \cite{Dihingia-etal2021} re-examined the underlying physical mechanisms for jets and winds generations and found that relativistic jets are driven by the BZ mechanism while the winds are ejected due to Blandford \& Payne (BP) mechanism. In a recent attempt, \cite{Aktar-etal2024} numerically investigated MHD accretion flow around spinning black hole and showed that mass outflow rate maintains positive correlation with the magnetic fields. All these works evidently suggest that the magnetic fields seem to play important role in generating jets/outflows from the accretion flow around the black holes. 

Meanwhile, there were attempts in the theoretical front to study the mass outflows around the black holes. Towards this, one of the earliest example is the advection dominated inflow-outflow solutions (ADIOS) around a Newtonian central object that involves the inward decrease of mass accretion rate resulting the mass loss throughout the disk \cite[]{Blandford-Begelman1999}. In parallel, efforts were also given to investigate the mass loss considering accretion shock-driven outflows around black holes \cite[]{Chakrabarti-1999,Das-etal-2001}. In these works, it was emphasized that outflows are emerged out due to strong coupling of accretion-ejection mechanism where advective accretion flow plays primitive role. In reality, during the course of accretion, rotating matter around black hole experiences centrifugal repulsion that eventually triggers the discontinuous transition of the flow variables in the form of shock waves. Such accretion solutions containing shocks are already studied in both hydrodynamic \cite[]{Fukue-1987, Chakrabarti-1989, Lu-etal-1999,Le-Becker-2005, Fukumara-Tsuruta-2004, Fukumura-Kazanas-2007,Dihingia-etal2018,Dihingia-etal2019a,Dihingia-etal2019b,Sen-etal-2022,Singh-Das2024} as well as magnetohydrodynamics \cite[]{Koide-etal1999,Takahashi-etal2002,Takahashi-etal2006,Takahashi-Takahashi2010,Sarkar-Das2016, Sarkar-etal2018,Das-Sarkar-2018} frameworks. Due to shock compression, convergent accretion flow becomes hot and dense in the post-shock region (equivalently post-shock corona, hereafter PSC) and therefore, PSC become puffed up resulting an effective boundary layer around the black hole. Because of this, an excess thermal gradient force is developed across the shock front which deflects a part of the accreting matter in the vertical direction to form bipolar outflows \cite[]{Molteni-etal-1994,Molteni-etal1996,Das-etal-2014,Okuda-2014,Okuda-Das-2015,Okuda-etal2019}. Such an appealing accretion-ejection mechanism successfully explains the disc-jet symbiosis involving increasing level of complexity around weakly rotating as well as rapidly rotating black holes \cite[]{Chakrabarti-1999,Das-etal-2001,chattopadhyay-Das-2007,das-chattopadhyay-2008,aktar-etal-2015,Kumar-Chattopadhyay2017,aktar-etal-2017,aktar-etal-2019}. However, all these works bear limitations as the effect of structured magnetic fields in estimating mass outflows were largely ignored although the presence of magnetic fields is ubiquitous in black hole systems. It is therefore timely to examine the role of structured magnetic fields in the generation of mass outflows from the magnetized accretion flow around black hole.

Being motivated with this, in this paper, we study the accretion-ejection mechanism considering a steady, relativistic, viscous, advective accretion flow threaded by the toroidal magnetic fields around rotating black hole. For simplicity, we adopt a recently developed effective potential \cite[]{Dihingia-etal-2018} that satisfactorily mimics the spacetime geometry around rotating black hole. With this, we self-consistently solve the coupled inflow-outflow governing equations and compute the mass outflow rate ($R_{\dot m}$) in terms of the inflow parameters (namely magnetic fields (plasma-$\beta$), accretion rates ($\dot m$) and viscosity ($\alpha_{\rm B}$)) and black hole spin ($a_{\rm k}$). We observe that mass outflow rate strongly depends on the magnetic fields as $R_{\dot m}$ increases when the magnetic activity is increased inside the disk. Further, we estimate the maximum mass outflow rate ($R^{\rm max}_{\dot m}$) by freely varying the model parameters of magnetized disk and find that rapidly rotating ($a_{\rm k}=0.99$) black hole yields higher $R^{\rm max}_{\dot m}$ than the weakly rotating ($a_{\rm k}\rightarrow 0$) black hole. Finally, using our model formalism, we attempt to explain the jet power observed from astrophysical black hole sources.

The plan of the paper is as follows: In the next Section, we present the assumptions and model description. In \S3, we discuss the obtained results in detail. In \S4, we explore our model formalism to explain the observed jet power from black hole sources. Finally, we end with summary and conclusion in \S 5.

\section{Assumptions and Governing equations}

We begin with the basic equations that govern an axisymmetric disk-jet system around a rotating black hole in the steady state. In particular, we assume that the accretion takes place along a disk geometry that remain confined around the black hole equatorial plane, whereas the jet geometry is described along the rotation axis of the black hole. Here, we adopt cylindrical polar coordinates ($x, \phi, z$), where black hole resides at its origin and disc extends along $z=0$ plane. We write all the equations in $M_{\rm BH}=G=c=1$ unit system, where $M_{\rm BH}$ is the mass of the black hole, $G$ is the gravitational constant and $c$ refers the speed of light. In this system, we express radial distance and angular momentum in units of $G M_{\rm BH}/c^2$ and $G M_{\rm BH}/c$, respectively. 

\subsection{Governing equations for accretion}

We consider a low angular momentum, relativistic, magnetized, viscous, advective accretion flow around a rotating black hole. In order to take care the effect of strong gravity, we adopt pseudo-potential \cite[]{Dihingia-etal-2018} that satisfactorily describes the spacetime geometry around rotating black hole. Following \cite{Hirose-etal-2006, Machida-etal-2006}, we consider the magnetic fields inside the disk are turbulent in nature and the azimuthal component of the magnetic field dominants over other components. With this, we write the azimuthally averaged magnetic field as $\left< \boldsymbol{B}\right> = \left< { B}_{\phi} \right> \hat \phi$, where `$\left < \hspace{1 mm} \right > $' implies azimuthal average and $B_{\phi}$ stands for azimuthal component of magnetic field \cite[]{Oda-etal-2007}.

In the steady state, the basic governing equations \cite[]{Sarkar-Das2016,Sarkar-etal2018} that describe the motion of the inflowing matter inside the accretion disc are as follows:

\noindent (a) The radial momentum equation: 
\begin{equation}
{v\frac{dv}{dx} + \frac{1}{\rho}\frac{dP}{dx} + \frac{d\Psi^{\rm eff}_{\rm e}}{dx} + \frac{\left<B_{\phi} ^2\right>}{4\pi x \rho} = 0},
\label{momenacc}
\end{equation}

\noindent (b) The Azimuthal momentum equation: 
\begin{equation}
v\frac{d\lambda}{dx}+\frac{1}{\Sigma x}\frac{d}{dx}(x^2T_{x\phi}) = 0,
\label{angular}
\end{equation}

\noindent (c) Mass flux conservation equation:
\begin{equation}
\dot{M} = 4 \pi  v\rho  h \sqrt{\Delta},
\label{massacc}
\end{equation}

\noindent (d) The entropy generation equation:
\begin{equation}
\Sigma v T \frac {ds}{dx}=\frac{hv}{\gamma-1}
\left(\frac{dp_{\rm gas}}{dx} -\frac{\gamma p_{\rm gas}}{\rho}\frac{d\rho}{dx}\right)=Q^- - Q^+,
\label{entropy}
\end{equation}

\noindent and (e) Radial advection of the toroidal magnetic flux:
\begin{equation}
\frac {\partial \left<B_{\phi}\right>\hat{\phi}}{\partial t} = {\bf \nabla} \times
\left({\vec{v}} \times \left<B_{\phi}\right>\hat{\phi} -{\frac{4\pi}{c}}\eta {\vec{j}}\right).
\label{induction}
\end{equation}
The variables $x$, $v$ and $\rho$ denote the radial distance, radial velocity and density of the inflow, respectively. The total isotropic pressure is given by $P=p_{\rm gas}+p_{\rm mag}$, where $p_{\rm gas}$ is the gas pressure and $p_{\rm mag}$ is the magnetic pressure. Here, the gas pressure is calculated as ${p_{ gas}} = R \rho T/\mu $, where $R$, $T$ and $\mu$ are the universal gas constant, local temperature of inflowing matter and mean molecular weight, respectively. For fully ionized hydrogen, we consider $\mu = 0.5$. The magnetic pressure inside the disk is calculated as ${ p_{\rm  mag}} = {\left< {B_{\phi}}^2 \right >} /{8 \pi}$, where $\left< {B_{\phi}}^2 \right >$ denotes the azimuthal average of the square of the toroidal
component of the magnetic field. We define plasma-$\beta ~( = p_{\rm gas}/p_{\rm mag})$ to express the total pressure as ${ P = p_{\rm gas}}(\beta + 1)/\beta$. The term $\Psi^{\rm eff}_{\rm e}$ denotes the effective potential on the disk equatorial plane and is given by,
\begin{equation}
\Psi^{\rm eff}_{\rm e} = \frac{1}{2}{\ln}\left[\frac{ x \Delta}{ a_{\rm k}^2(x +2) -  4a_{\rm k} \lambda +  x^{3} - \lambda^{2}(x-2)}\right]
\label{potential1},
\end{equation}
where $\lambda$ is the local specific angular momentum (hereafter angular momentum), $a_{\rm k}$ is the Kerr parameter that measures the spin of the black hole, and $\Delta =  {x^2 - 2x + a_{\rm k}^2}$. The subscript `$\rm e$' refers to the quantity measured on the equatorial plane.
In equation (\ref{angular}), we consider the vertically integrated total stress which is dominated by the $x \phi$ component of the Maxwell stress $T_{x \phi}$ over other components. Following the simulation work of \cite{Machida-etal-2006}, we estimate $T_{ x\phi}$ for an advective flow possessing significant radial velocity as \cite[]{chakrabarti-1996}),
\begin{equation}
T_{x\phi} = \frac{<B_{x}B_{\phi}>}{4\pi}h = -\alpha_{\rm B}(W + \Sigma v^2),
\label{viscosity}
\end{equation}
where $W$ and $\Sigma$ denote the vertically integrated pressure and density of the inflow \cite[]{Matsumoto-etal-1984}, and $\alpha_{\rm B}$ (ratio of Maxwell stress to the total pressure) is the constant of proportionality. In this work, following the seminal work of \cite{Shakura-Sunyaev-1973}, we refer $\alpha_{\rm B}$ as viscosity parameter.
When the inflow velocity becomes negligible as in the case of standard Keplerian disk, equation (\ref{viscosity}) reduces to the original prescription of the `$\alpha$-model' \cite[]{Shakura-Sunyaev-1973}. In equation (\ref{massacc}), the mass accretion rate is denoted by ${\dot M}$ and $h$ refers the local half-thickness of the disk. Following \cite{Riffert-Herold-1995, peitz-Appl-1997}, we calculate $h$ as,
\begin{equation}
h = \sqrt{\frac{ P x^3}{\rho \mathcal{F}}}; \quad \mathcal{F} = \frac{1}{1-\lambda \Omega} \times \frac{ (x^2 + a_k^2)^2 + 2 \Delta  a_{\rm k}^2}{ (x^2 + a_{\rm k}^2)^2 - 2 \Delta  a_{\rm k}^2},
\label{height}
\end{equation}
where $\Omega ~[= (2 a_{\rm k} + \lambda (x-2))/(a^2_{\rm k}(x + 2) - 2 a_{\rm k} \lambda + x^3)]$ is the angular velocity of the flow. We define the sound speed of the inflow as $a = \sqrt{{\gamma  P}/\rho}$, where $\gamma$ is the adiabatic index. Here, we assume $\gamma$ to remain constant and choose a canonical value of $\gamma = 4/3$ all throughout unless otherwise stated. In equation (\ref{entropy}), $s$ and $T$ denote the specific entropy and the local temperature of the inflow. Here, $Q^{+}$ and $Q^{-}$ are the vertically integrated heating and cooling rates, respectively. In reality, simulation studies reveal that flow is heated due to the  thermalization of magnetic energy through the magnetic reconnection process \cite[]{Hirose-etal-2006, Machida-etal-2006}, and hence, the heating rate is expressed as,
\begin{equation}
Q^{+} = \frac{<B_{x}B_{\phi}>}{4\pi} x h \frac{d\Omega}{dx} = 
-\alpha_{\rm B}(W + \Sigma v^2) x \frac{d\Omega}{dx}.
\label{heatgain}
\end{equation}
On the contrary, the cooling of the accreting matter is governed by various radiative processes, namely bremsstrahlung, synchrotron, Comptonization of bremsstrahlung and synchrotron photons. However, in this work, since we deal with the magnetized accretion flow, it is evident that the synchrotron process would become effective to cool the flow. Accordingly, we obtain the cooling rate due to synchrotron radiation \cite[]{Shapiro-Teukolsky-1983} which is given by,
\begin{eqnarray}
Q^{-} &=& \frac{ S a^5  \rho  h }{ v} \sqrt{\frac{\mathcal{F}}{ x^3 \Delta}}\frac{\beta^2}{(1+\beta)^3};\\
S &=& 1.4827 \times 10^{18} \frac{\dot{m} \mu^2 e^4}{ m_{e}^3\gamma^{5/2}}\frac{1}{\ GM_{\odot}c^3},\nonumber
\label{heatcool}
\end{eqnarray} 
where $m_e$ and $e$ specify the mass and charge of electron, respectively. Following the work of \cite{Chattopadhyay-Chakrabarti-2002}, we estimate the electron temperature for a single temperature flow as $T_{e} = \sqrt{m_e/m_p} T_p$ ignoring any coupling between the ions and electrons, where $m_p$ and $T_p~(=T)$ are the mass and temperature of ion, respectively. We express the accretion rate as $ \dot{m} = \dot{M}/\dot{M}_{\rm Edd}$, where $\dot{M}_{\rm Edd}~\left(=1.39 \times 10^{17} M_{\rm BH}/M_{\odot} {\rm ~g~s}^{-1}\right)$ represent the Eddington accretion rate. It may be noted that in this work, we ignore bremsstrahlung cooling since it is regarded as a very inefficient cooling process \cite[]{Chattopadhyay-Chakrabarti-2002,Das-2007}. We also neglect inverse-Comptonization as it requires two temperature accretion solutions and obtaining such solutions is beyond the scope of the present work. Finally, the advection rate of toroidal magnetic field is described using the induction equation and its azimuthal averaged form is presented in equation (\ref{induction}), where $\vec {v}$ refers the velocity vector, $\eta$ is the resistivity and $\vec{j} = c(\nabla \times {\left< B_{\phi}\right>} \hat{\phi})/4\pi$ denotes the current density. In general, as the Reynolds number remains very large in an accretion disk due to its large extent, we neglect the magnetic-diffusion term. In addition, we also ignore the dynamo term. With this, we obtain the vertically averaged resultant equation considering the fact that the azimuthally averaged toroidal magnetic fields vanish at the disc surface. Accordingly, we obtain the advection rate of the toroidal magnetic flux
as \cite[]{Oda-etal-2007},
\begin{equation}
\dot{\Phi} = - \sqrt{4\pi}v h {B}_{0} (x),
\label{magflux}
\end{equation}
where $B_{0}(x)$ represents the azimuthally averaged toroidal magnetic field confined
in the disc equatorial plane and given by,
\begin{align*}
{B}_{0} (x) & = \langle {B}_{\phi} \rangle \left(x; z = 0\right),  \\
& = 2^{5/4}{\pi}^{1/4}(R T/\mu)^{1/2}{\Sigma}^{1/2}h^{-1/2}{\beta}^{-1/2}.
\label{magfield}
\end{align*}
In reality, $\dot{\Phi}$ is not a conserved quantity, rather expected to vary with $x$ in presence of the dynamo and the magnetic-diffusion terms. However, in the quasi steady state, the global 3D MHD simulation \cite[]{Machida-etal-2006} suggests that $\dot{\Phi} \propto 1/ x$. Indeed, the explicit computation involving both dynamo and the magnetic-diffusion terms are very much complex and tedious, and hence, we introduce a parameter $\zeta$ to adopt a parametric relation \cite[]{Oda-etal-2007} as,
\begin{equation}
\dot{\Phi}\left(x; \zeta, \dot{M}\right) \equiv \dot{\Phi}_{\rm edge} 
\left(\frac{x}{x_{\rm edge}} \right)^{-\zeta},
\label{fluxvary}
\end{equation}
where $\dot{\Phi}_{\rm edge}$ is the advection rate of the toroidal magnetic flux at the outer edge of the disk ($ x_{\rm edge}$). In this work, for the purpose of representation, we choose $\zeta = 1$ in the subsequent analysis unless stated otherwise.

Using equations (\ref{momenacc}), (\ref{angular}), (\ref{massacc}), (\ref{entropy}), (\ref{magflux}) and (\ref{fluxvary}), and after some algebra, we obtain the wind equation which is given by,
\begin{equation}
\frac{dv}{dx} = \frac{{\cal N}(x, v, a, \lambda, \beta)}{{\cal D}(x, v, a, \lambda, \beta)},
\label{dvdx}
\end{equation} 
where $\cal N$ and $\cal D$ are the numerator and denominator and their explicit expressions are given in Appendix A. Using $dv/dx$ (equation (\ref{dvdx})), we express the derivatives of the sound speed ($a$), angular momentum ($\lambda$) and plasma-$\beta$ as
\begin{equation}
\frac{ da}{ dx} =  a_{11} + a_{12} \frac{dv}{dx},
\label{dadx}
\end{equation}
\begin{equation}
\frac{ d\lambda}{ dx} =  \lambda_{11} + \lambda_{12} \frac{ dv}{ dx},
\label{dldx}
\end{equation}
\begin{equation}
\frac{ d\beta}{ dx} =  \beta_{11} + \beta_{12} \frac{ dv}{ dx},
\label{dbetadx}
\end{equation}
where the coefficients $a_{11}, a_{12}$, $\lambda_{11}$, $\lambda_{12}$, $\beta_{11}$ and $\beta_{12}$ are the explicit functions of flow variables which are given in Appendix A.

The advective accretion flow around black holes must be transonic in order to satisfy the inner boundary conditions imposed by the event horizon. In reality, the flow starts accreting with a subsonic velocity ($v < a$) from the outer edge ($x_{\rm edge}$) of the disk and gradually gains radial velocity as it moves inward. During the course of accretion, flow is also compressed causing the increase of density, temperature and sound speed with decreasing radii. Ultimately, flow crosses the event horizon with velocity equivalent to the speed of light implying that the flow is supersonic close to the BH. This evidently indicates that accretion flow must smoothly pass through the critical point ($x_{\rm c}$) where the sonic transition occurs. At the critical point, both $\cal N$ and $\cal D$ of equation (\ref{dvdx}) vanish simultaneously and hence, radial velocity gradient takes to form $(dv/dx)_{\rm c}=0/0$. Since the flow velocity remains smooth along the streamline, $dv/dx$ must be real and finite all throughout. Hence, we calculate $dv/dx|_{\rm c}$ by applying the l$'$H\^{o}pital's rule as $(dv/dx)_{x_{\rm c}} = [(d \mathcal{N}/dx)/(d\mathcal{D}/dx)]_{x_{\rm c}}$. For a physically acceptable transonic solution, flow must contain at least one saddle-type sonic point \cite[and references therein]{Das-2007}. Depending of the input parameters, flow may possess multiple critical points as well which is one of the necessary condition for shock formation \cite[]{Chakrabarti1990}. In general, inner critical point ($x_{\rm in}$) is formed close to the horizon, whereas outer sonic point ($x_{\rm out}$) is resided far away from the black hole.

In order to obtain the global magnetized transonic accretion solution around black hole, we  simultaneously solve the equations (\ref{dvdx}-\ref{dbetadx}) for a set of flow parameters \cite[and references therein]{Sarkar-etal2018}. In this analysis, we treat $\alpha_{\rm B}$, $\dot m$ and $\gamma$ as global flow parameters, whereas the boundary values of $\lambda$ and $\beta$ at $x_{\rm c}$ are required as local flow parameters to solve these equations. In addition, we need $a_{\rm k}$ value as well to specify the spinning nature of the black hole. With all these input parameters, we integrate equations (\ref{dvdx}-\ref{dbetadx}) starting from $x_{\rm c}$ inwards up to just outside the BH horizon and then outwards up to $x_{\rm edge}$ to obtain a global accretion solution.

\subsection{Governing equations for outflows}
\label{eqn-outflow}

We consider that the outflow is originated from the accretion disc and the outflow geometry is oriented along the rotational axis of the black hole. Since a part of the accreting matter is emerged out in the form of outflow, it is expected that outflows are tenuous in nature. Because of this, we ignore the effect of viscosity in outflows as the differential rotation of the outflowing matter is likely to be negligibly small. Moreover, as the toroidal component of the magnetic field is considered as the dominant one and the outflows are streamed along the axial direction, we neglect the effect of magnetic fields while describing outflows for simplicity. What is more is that we consider the outflow to obey the polytropic equation of state as $P_{\rm j} = K_{\rm j} \rho^\gamma_{\rm j}$, where the subscript `j' stands for outflow variables and $K_{\rm j}$ refers the measure of entropy of the outflow. Based on the above assumptions, we have the following governing equations that describe the outflow motion as,

\noindent (i) Energy conservation equation:
\begin{equation}
	\mathcal{E}_{\rm j} = \frac{1}{2}v^2_{\rm j} + \frac{a^2_{\rm j}}{\gamma -1} + \Psi^{\rm eff},
	\label{outeng}
\end{equation}
where $\mathcal{E}_{\rm j}$, $v_{\rm j}$ and $a_{\rm j}$ are the energy, velocity and sound speed of the outflowing matter, respectively. Here, $\Psi^{\rm eff}$ denotes the effective potential at the off equatorial plane around the rotating black hole.

\begin{figure}
	\includegraphics[width=\columnwidth]{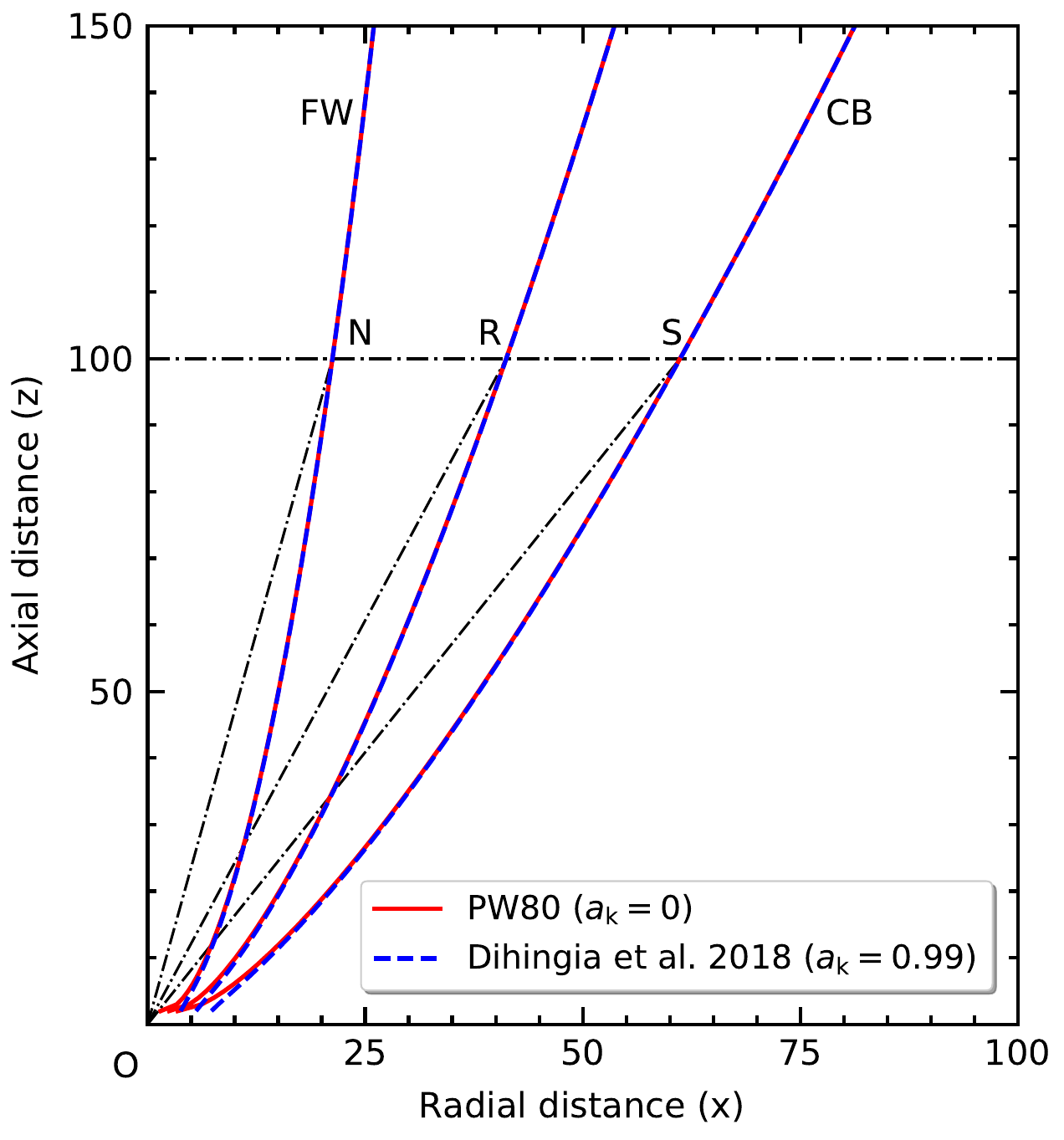}
	\caption{Comparison of jet geometries obtained from different  pseudo-potentials. Dashed (blue) curve denote result obtained for $a_{\rm k} = 0.99$ \citep{Dihingia-etal-2018}, whereas solid (red) curve corresponds to $a_{\rm k} = 0$ \citep[PW80]{Paczynsky-Wita-1980}. Here, ON (= $r_{\rm FW}$) denotes spherical radius of funnel wall (FW), OR (= $ r_{\rm j}$) is the jet spherical radius and OS (= $r_{\rm CB}$) is spherical radius of centrifugal barrier (CB). We choose $\lambda_{\rm j} = 3.0$. See the text for details. 
		}
	\label{geom}
\end{figure}

\noindent (ii) Mass conservation equation:
\begin{equation}
	\dot{M}_{\rm out} = \rho_{\rm j} v_{\rm j} \mathcal{A}_{\rm j},
	\label{outmass}
\end{equation}
where $\dot{M}_{\rm out}$ is the mass outflowing rate and $\mathcal{A}_{\rm j}$ is the area function of the outflow. Following \cite{Molteni-etal1996}, we calculate $\mathcal{A}_{\rm j}$ considering the fact that the outflowing matter are ejected out between the surfaces of funnel wall ($\rm FW$) and centrifugal barrier ($\rm CB$) \cite[]{chattopadhyay-Das-2007,das-chattopadhyay-2008}. The FW refers the pressure minimum surface defined by the null effective potential as $\Psi^{\rm eff}\vert_{ r_{\rm FW}}$ = 0, whereas the CB is identified as the pressure maxima surface defined as $\left(\frac{ d\Psi^{\rm eff}}{ dx}\right)_{r_{\rm CB}} = 0$. Similar to accretion process, we have the expression of the effective potential as \cite[]{Dihingia-etal-2018},
\begin{equation}
	\begin{split}
		\Psi^{\rm eff} & = \frac{1}{2} \ln \, \frac{x^2_{\rm j} (2 {\mathcal Y} r_{\rm j} - 4 a^2_{\rm k}x^2_{\rm j} - {\mathcal Y} {\mathcal Z})}{{\mathcal Z} [- {\mathcal Y} x^2_{\rm j} + 4 a_{\rm k} \lambda_{\rm j}r_{\rm j}x^2_{\rm j} + \lambda^2_{\rm j}r^2_{\rm j}({\mathcal Z} - 2 r_{\rm j})]},
	\end{split}
	\label{pot-off}
\end{equation}
where $r_{\rm j}~\left(=\sqrt{x^2_{\rm j} + z^2_{\rm j}}\right)$ is the spherical radius of outflow, $x_{\rm j} = (x_{\rm CB} + x_{\rm FW})/2$, $z_{\rm j} = z_{\rm FW} = z_{\rm CB}$ (see Fig. \ref{geom}), ${\mathcal Y} =  (r^2_{\rm j} + a^2_{\rm k})^2- \Delta  a_{\rm k}^2 (x_{\rm j}/r_{\rm j})^2$ and ${\mathcal Z} = { r^2_{\rm j} + a^2_{\rm k}\Big(1 - (x_{\rm j}/r_{\rm j})^2\Big)}$, respectively. 

In Fig. \ref{geom}, we depict the schematic diagram of the outflow geometries for rotating black holes that are calculated numerically for $\lambda_{\rm j}= 3.0$. Here, the region bounded by the dashed (blue) curves are for extreme spin value $a_{\rm k} = 0.99$. We compare this outflow geometry with the same obtained from the well known pseudo-Newtonian potential \cite[]{Paczynsky-Wita-1980} (PW80) for stationary black hole ($a_{\rm k}=0$) and plotted using solid curves (red). From the figure, it is evident that outflow geometries obtained for both rotating and stationary black holes remain largely indistinguishable particularly for $x \gtrsim 10r_g$. This happens because the effect of black hole spin on the spacetime geometry rapidly reduces as radial distance increases. Since the pseudo-Newtonian potential provides the analytical forms of both FW and CB, it is straight forward to calculate the area function (${\cal A}$). Hence, in this work, we adopt the outflow geometry of stationary black hole to avoid the rigorous numerical calculations in obtaining the outflow geometry.

Similar to accretion, we carry out the critical point analysis to solve the jet equations \cite[and references therein]{Das-etal-2001a}. Using equations (\ref{outeng}) and (\ref{outmass}), we get the critical point condition for jet \cite[]{Kumar-Chattopadhyay-etal-2013} as,
\begin{equation}
	v_{\rm jc} = a_{\rm jc} = \sqrt{ \left(\frac{d \Psi^{\rm eff}}{dr}\right)_{r_{\rm jc}}
		\left[\frac{1}{\mathcal A_{\rm j}}\Big(\frac{ \mathcal{A}_{\rm j}}{dr}\Big)_{r_{\rm jc}}\right]^{-1} },
\label{jetcritical}
\end{equation}
where $r_{\rm jc}$ denotes the jet critical point, and $v_{\rm jc}$ and $a_{\rm jc}$ are the matter speed and sound speed at $r_{\rm jc}$. Here, $r=r_{\rm CB}$. In general, since jet streamline and jet area vector remain misaligned, we incorporate the projection factor $\sqrt{1 + ( dx_{\rm j}/dz_{\rm j})^2}$ while  calculating the jet area function as $\mathcal{A}_{\rm j} = 2 \pi (x^2_{\rm CB} - x^2_{\rm FW})/\sqrt{1 + ( dx_{\rm j}/dz_{\rm j})^2}$ \cite[]{Kumar-Chattopadhyay-etal-2013}. Using the sonic point condition, we solve the outflow equations (\ref{outeng} and \ref{outmass}) and obtain outflow/jet solution uniquely for a given set of $\mathcal{E}_{\rm j}$ and $\lambda_{\rm j}$. In order to obtain the self-consistent accretion-ejection solution, we couple the accretion and outflow solutions in the next section and subsequently, we examine the outflow properties in terms of the inflow parameters ($\beta$, $\alpha_{\rm B}$ and $\dot m$).

\subsection{Disk-jet connection} 

In reality, the rotating matter experiences centrifugal repulsion while accreting towards the black hole. When centrifugal force is comparable to the gravity, matter starts accumulating in the vicinity of the black hole. Because of this, a puffy torus like structure is formed that acts like an effective boundary layer around black holes (equivalently post-shock corona, hereafter PSC), triggering the shock transition ($x_s$). Indeed, the fast moving pre-shock flow converts its kinetic energy to heat up the post-shock region. This excess thermal gradient force leads to eject a part of the accreting matter as bipolar outflow/jet in the vertical direction. As discussed in \S\ref{eqn-outflow}, we assume the outflow to be guided by the CB and FW surfaces \cite[]{Molteni-etal1996}. Since the accretion and ejection processes are coupled via PSC, in this work, we solve the inflow-outflow equations self-consistently using the shock conditions. In presence of mass loss, the shock conditions \cite[]{Landau-Lifshitz-1959,das-chattopadhyay-2008,Sarkar-etal2018} for vertically averaged accretion flow are given as \noindent(a) the energy flux conservation: $\mathcal{E}_{+} = \mathcal{E}_{-}$, (b) the mass flux conservation $\dot{M}_{+} = \dot{M}_{-} - \dot{M}_{\rm out} = \dot{M}_{-}(1-R_{\dot{m}})$, (c) the momentum flux conservation: $W_{+} + \Sigma_{+} v^2_{+} = 	W_{-} + \Sigma_{-} v^2_{-}$ and (d) the magnetic flux conservation: $\dot{\Phi}_{+} = \dot{\Phi}_{-}$, respectively. Here, the suffix `$-~(+)$' denotes pre(post)-shock quantities across the shock front, $\mathcal{E} ~[=  v^2/2 + a^2/(\gamma -1) + \left < B^2_{\phi} \right >/(4\pi \rho) + \Psi^{\rm e}_{\rm eff}]$ refers the local inflow energy, and $R_{\dot{m}} ~(= \dot{M}_{\rm out}/\dot{M}_{-})$ is the mass outflow rate. In the present formalism, as the outflow/jet must originate from the PSC, we assume that it emerges with same local $\mathcal{E}$, $\lambda$ and $\rho$ of the post-shock flow immediately after the shock. Accordingly, we have the outflow/jet variables at the jet base as $\mathcal{E}_{\rm j} =\mathcal{E}_{+}$, $\lambda_{\rm j} = \lambda_{+}$ and $\rho_{\rm j} = \rho_{+}$. Using these boundary values, we numerically solve jet equations (\ref{outeng} and \ref{outmass}) starting from the critical point and obtain the outflow variables ($v_{\rm jb}, a_{\rm jb}, {\mathcal A}_{\rm jb}$) at the jet base ($x_s$). Subsequently, using equations (\ref{massacc} and \ref{outmass}), we compute the mass outflow rate ($R_{\dot{m}}$) as,
\begin{equation}
	  R_{\dot{m}} = \frac{v_{\rm jb} \mathcal A_{\rm jb}}{4 \pi a_{+} v_{-}}\left(\frac{\Sigma_{+}}{\Sigma_{-}}\right)\left( \frac{\gamma {\cal F}}{\Delta x^3_s}\right)^{1/2}.
	 \label{rmdot}
\end{equation}

In order to obtain the self-consistent accretion-ejection solutions, we solve the coupled inflow-outflow equations simultaneously adopting the following approach \cite[]{chattopadhyay-Das-2007,das-chattopadhyay-2008}. To begin with, we consider $R_{\dot{m}} = 0$, and compute the virtual shock location ($x^{*}_{s}$) by supplying the model input parameters (see \S 2.1). Once $x^{*}_{s}$ is known, we assign the jet variables ($i.e.$, $\mathcal{E}_{\rm j}$, $\lambda_{\rm j}$ and $\rho_{\rm j}$) to solve the jet equations and calculate $R_{\dot{m}}$ by using equation (\ref{rmdot}). We use this value of $R_{\dot{m}}$ to calculate the updated shock location. We continue this successive iteration until the shock location converges, and accordingly, we obtain the mass outflow rate $R_{\dot m}$.

\section{Results}

In this work, we are interested to examine the mass loss from magnetized accretion disk. Towards this, in the subsequent sections, we mainly focus on the self-consistent accretion-ejection solutions around rotating black hole. 

\subsection{Global inflow-outflow solution (GIOS)}

\begin{figure}
	\begin{center}
		\includegraphics[width=\columnwidth]{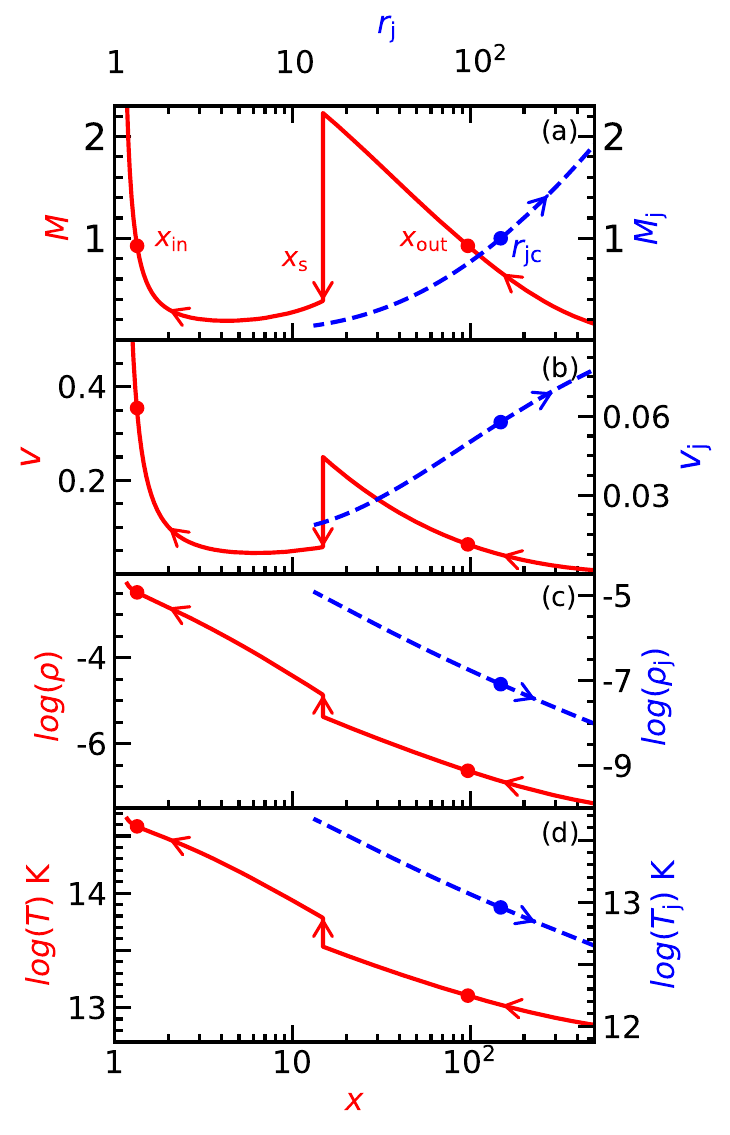}
	\end{center}
	\caption{Typical example of GIOS, where the variation of inflow Mach numbers $M$ (outflow Mach number $M_{\rm j}$), inflow velocity $v$ (outflow velocity $v_{\rm j}$), inflow density $\rho$ (outflow density $\rho_{\rm j}$) and inflow temperature $T$ (outflow temperature $T_{\rm j}$) are depicted as function of $x$ ($r_{\rm j}$). Here, we choose the global parameters as $\dot{m}=0.001$, $\alpha_{\rm B}=0.01$, and $a_{\rm k} = 0.99$, whereas the local inflow parameters at $x_{\rm in} = 1.3375$ is fixed as $\lambda_{\rm in} = 2.05$, ${\cal E}_{\rm in} = 4.322 \times 10^{-3}$, and $\beta_{\rm in} = 300$, respectively. Solid (red) and dashed (blue) curves denote the inflow and outflow solutions and filled circles refer the critical points. Arrows indicate the direction of flow motion and vertical arrow shows the shock location. See the text for details.
	}
	\label{soltn}
\end{figure}

In Fig. \ref{soltn}, we illustrate a typical example of the coupled global inflow-outflow solution (GIOS), where both inflow and outflow variables are plotted with radial coordinates ($x$ and $r_{\rm j}$). In doing so, we use lower $x$-axis and left $y$-axis to demonstrate the inflow variables, whereas upper $x$-axis and right $y$-axis are used to display the jet variables. Here, we choose $\dot{m}=0.001$, $\alpha_{\rm B}=0.01$ and $a_{\rm k} = 0.99$, and supply the local inflow parameters at the inner critical point $x_{\rm in} = 1.3375$ as $\lambda_{\rm in} = 2.05$, ${\cal E}_{\rm in} = 4.322 \times 10^{-3}$, and $\beta_{\rm in} = 300$, respectively. With this, we self-consistently obtained GIOS and depict the variation of inflow Mach number ($M$) and outflow Mach number ($M_{\rm j}$) using solid (red) and dashed (blue) curves in Fig. \ref{soltn}a. During the course of accretion, subsonic inflowing matter from the outer edge of the disk ($x_{\rm edge} = 500$) becomes supersonic after crossing the outer critical point at $x_{\rm out} = 96.64$ and continues to proceed towards the black hole. Meanwhile, supersonic matter starts experiencing centrifugal repulsion that triggers the discontinuous transition of the inflow variables in the form of shock waves at $x_s = 14.826$ in presence of mass loss. Indeed, after the shock, a part of the inflowing matter is deflected to form outflows and the remaining matter enters in to the black hole supersonically after crossing the inner critical point at $x_{\rm in} = 1.3375$. We find that GOIS renders the energy and angular momentum of the outflow as ${\cal E}_{\rm j} = 6.179 \times 10^{-3}$ and $\lambda_{\rm j} = 2.06$ that provides the jet critical point at $r_{\rm jc} = 147.83$, and ultimately, we obtain the mass outflow rate as $R_{\dot{m}} = 0.11$. In the figure, filled circles refer the location of the critical points, arrows indicate the over all direction of the flow motion and the vertical arrow represents the shock radius ($x_s$). In Fig. \ref{soltn}b, we present the variation of $v$ and $v_{\rm j}$ corresponding to the solution presented in Fig. \ref{soltn}a. We observe that $v$ drops abruptly across the shock front, however, it again increases as the flow proceeds further while entering in to the black hole. On the other hand, $v_{\rm j}$ is seen to increase as outflow moves away from the BH and it tends to achieve a terminal speed exceeding $0.075 c$ at $r_{\rm j} = 500$. Next, in Fig. \ref{soltn}c, we show the profiles of $\rho$ and $\rho_{\rm j}$ for the same solution as shown in Fig. \ref{soltn}a. It is evident that convergent accreting matter experiences shock compression and hence, $\rho$ jumps up during the shock transition. We also observe that $\rho_{\rm j}$ decrease with $r_{\rm j}$. Finally, we demonstrate the temperature ($T$ and $T_{\rm j}$) variations of inflow and outflow in Fig. \ref{soltn}d. Across the shock, supersonic matter jumps in to the subsonic branch, and because of this, kinetic energy is converted to thermal energy yielding the increase of temperature at the PSC. As the outflow is launched from the PSC, outflow temperature is high at the jet base, however, $T_{\rm j}$ decreases as jet moves away from the black hole. 

\subsection{Effect of magnetized accretion on outflow rate}

\begin{figure}
	\includegraphics[width=\columnwidth]{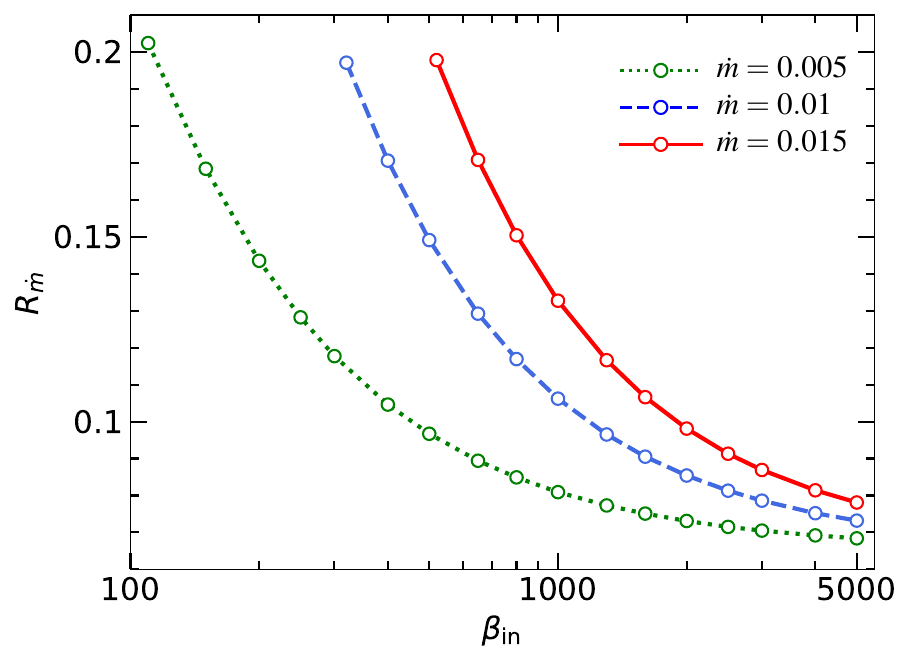}
	\caption{Variation of mass outflow rate ($R_{\dot m}$) as function of $\beta_{\rm in}$ for different accretion rate ($\dot m$). Open circles joined with dotted, dashed and solid lines are for $\dot{m} = 0.005$ (green), $0.01$ (blue) and $0.015$ (red), respectively. Here, $a_{\rm k} = 0.9$ and $\alpha_{\rm B} = 0.01$. The inflow parameters fixed at inner critical points are $\mathcal{E}_{\rm in} = 5.8 \times 10^{-3}$ and $\lambda_{\rm in} = 2.34$. See the text for details.}
	\label{mdot}
\end{figure}

In Fig. \ref{mdot}, we depict the variation of $ R_{\dot{m}}$ as a function of $\beta_{\rm in}$ for a set of $\dot{m}$ values. Here, we choose the global parameters as $\rm a_{k} = 0.9$ and $\alpha_{\rm B} = 0.01$, and fix the inflow parameters at the inner critical point ($x_{\rm in}$) as  $\mathcal{E}_{\rm in} = 5.8 \times 10^{-3}$ and $\lambda_{\rm in} = 2.34$. The obtained results plotted using dotted (green), dashed (blue) and solid (red) curves correspond to $\dot{m} = 0.005, 0.01$ and $0.015$, respectively. We observe that for a fixed $\dot{m}$, the outflow rate ($R_{\dot m}$) increases as $\beta_{\rm in}$ decreases. This finding suggests that as the disc becomes more magnetized, the possibility of mass loss from the disc increases. Indeed, when magnetic fields are increased, the cooling becomes more efficient resulting the decrease of energy ${\cal E}(x)$ as flow accretes. Accordingly, for lower $\beta_{\rm in}$, flow starts accreting from the outer part of the disk with higher energy just to maintain identical ${\cal E}_{\rm in}$ and hence, flow possesses higher energy at the shock. Such an increase of energy at the PSC drives more matter from the post-shock region as outflows, resulting an increase in $R_{\dot{m}}$.  We also notice that for a given $\beta_{\rm in}$, $R_{\dot m}$ increases with the increase in $\dot{m}$. This happens because, higher $\dot m$ enhances the cooling of the accreting matter that yields the local energy of the flow higher at PSC for fixed ${\cal E}_{\rm in}$. This eventually causes the excess driving to deflect more matter at PSC in the form of outflow.

\begin{figure}
	\includegraphics[width=\columnwidth]{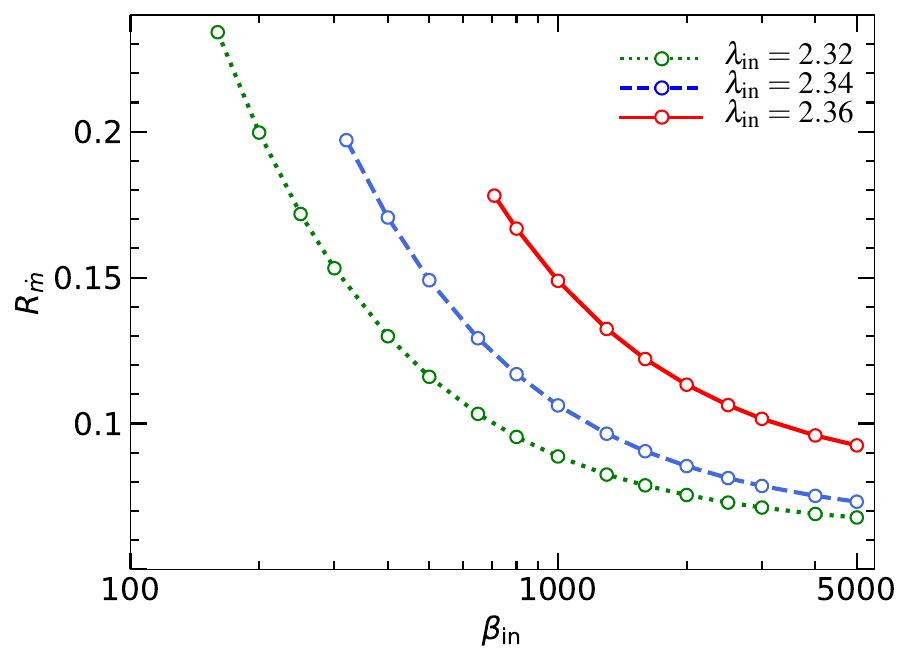}
	\caption{Variation of mass outflow rate ($R_{\dot m}$) with $\beta_{\rm in}$ for different $\lambda_{\rm in}$. Open circles joined with dotted, dashed and solid lines are for $\lambda_{\rm in}= 2.32$ (green), $2.34$ (blue) and $2.36$ (red), respectively. Here, we choose $\dot{m} = 0.01$, $a_{\rm k} = 0.9$, $\alpha_{\rm B} = 0.01$ and $\mathcal{E}_{\rm in} = 5.8 \times 10^{-3}$. See the text for details.}
	\label{lambda}
\end{figure}

Next, we examine how $R_{\dot{ m}}$ varies with $\beta_{\rm in}$ for a set of different inflow angular momentum $\lambda_{\rm in}$ fixed at $x_{\rm in}$. Here, we choose input parameters as $\mathcal{E}_{\rm in} = 5.8 \times 10^{-3}$, $\dot{m} = 0.01$, $a_{\rm k} = 0.9$ and $\alpha_{\rm B} = 0.01$. The results are depicted in Fig. \ref{lambda} where dotted (green), dashed (blue) and solid (red) curves are for $\lambda_{\rm in} = 2.32, 2.34$ and $2.36$, respectively. We observe that for a given $\beta_{\rm in}$, higher angular momentum flow provides higher $R_{\dot{ m}}$. This is not surprising as the higher $\lambda_{\rm in}$ enhances the centrifugal repulsion that pushes the shock front away from the horizon resulting expanded PSC. Since PSC basically acts as the jet base, this leads to the net outflow rate higher. Moreover, we also find that $R_{\dot m}$ increases with $\beta_{\rm in}$ for fixed $\lambda_{\rm in}$ which is very much expected (see Fig. \ref{mdot}). With this, we emphasize that  both ${\cal E}_{\rm in}$ and $\lambda_{\rm in}$ play crucial role in determining the mass outflow rate ($R_{\dot{ m}}$) from a magnetized accretion disk.

\begin{figure}
	\includegraphics[width=\columnwidth]{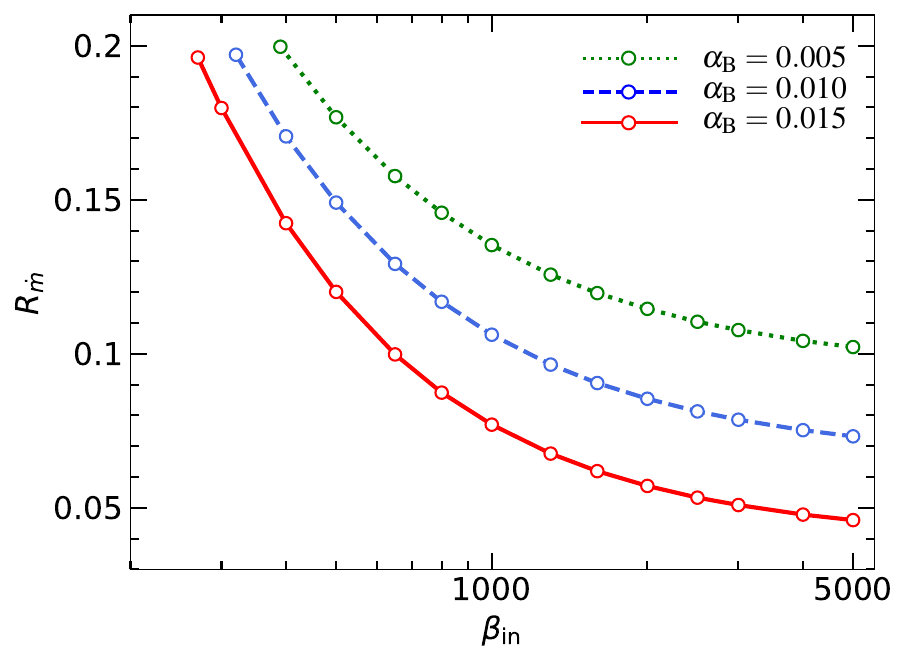}
	\caption{ $R_{\dot{m}}$ variation with  $\beta_{\rm in}$ for a set of $\alpha_{\rm B}$ values. Open circles joined with dotted, dashed and solid lines are for $\alpha_{\rm B} = 0.005$ (green), $0.01$ (blue) and $0.015$ (red), respectively. Here, $a_{\rm k} = 0.9$ and $\dot{m} = 0.01$. The inflow parameters at inner critical point are chosen as $\mathcal{E}_{\rm in} = 5.8 \times 10^{-3}$ and $\lambda_{\rm in}$ = 2.34. See the text for details.}
	\label{alpha}
\end{figure}

In Fig. \ref{alpha}, we compare the variation of $R_{\dot{ m}}$ with $\beta_{\rm in}$ for different values viscosity parameters ($\alpha_{\rm B}$). In obtaining the results, we choose $\mathcal{E}_{\rm in} = 5.8 \times 10^{-3}$, $\lambda_{\rm in}$ = 2.34, $a_{\rm k} = 0.9$ and $\dot{m} = 0.01$. In the figure, dotted (green), dashed (blue) and solid (red) curves represent results obtained for $\alpha_{\rm B} = 0.005, 0.01$ and $0.015$, respectively. We observe that for a fixed $\beta_{\rm in}$, the increase of $\alpha_{\rm B}$ leads to the decrease of $R_{\dot{m}}$. It may be noted that viscosity plays dual role for flows accreting on to a black hole. In one hand, $\alpha_{\rm B}$ transports angular momentum outwards, while in other hand, flow gains energy due to viscous dissipation as it accretes towards the black hole. Therefore, the combined effects of viscosity operates inherently in generating outflow from the accretion disc. We notice that higher $\alpha_{\rm B}$ causes a moderate increase of angular momentum at $x_s$, whereas sharp decrease of flow energy $\mathcal{E}(x_s)$ is observed (as ${\cal E}_{\rm in}$ is kept fixed at $x_{\rm in}$). This apparently weakens the jet driving at the PSC yielding the subsequent decrease of $R_{\dot{m}}$ for flows with higher viscosity. Furthermore, when $\alpha_{\rm B}$ is fixed, $R_{\dot{m}}$ increases monotonically with the decrease of $\beta_{\rm in}$. This is not surprising because for a convergent flow of fixed inner boundary, the energy at the PSC increases as cooling is increased that results the inevitable increase of $R_{\dot{m}}$.

\begin{figure}
	\includegraphics[width=\columnwidth]{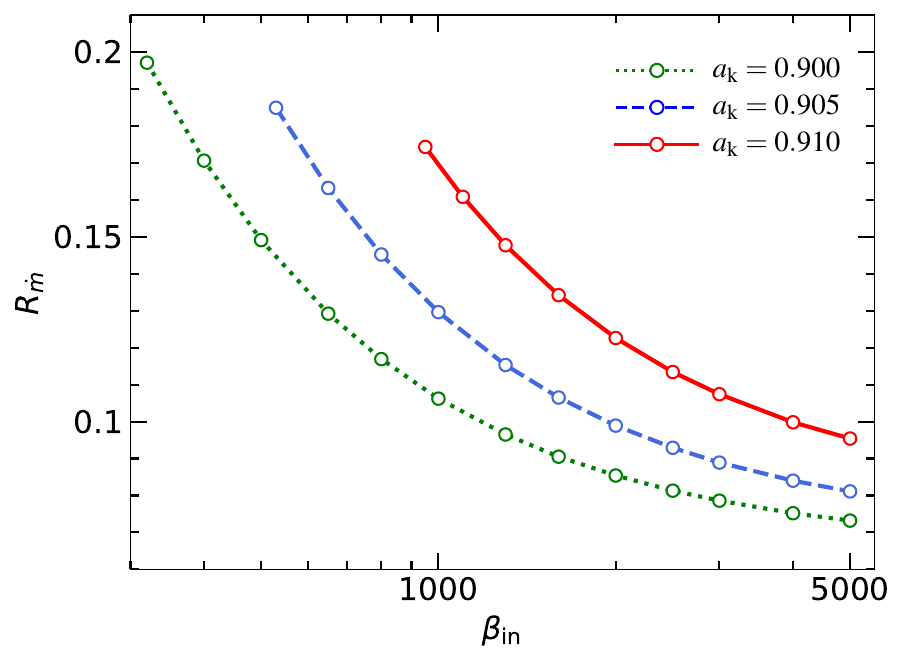}
	\caption{Variation of $R_{\dot m}$ with $\beta_{\rm in}$ for different black hole spin ($a_{\rm k}$). Open circles joined with dotted, dashed and solid lines are for $a_{\rm k} = 0.9$ (green), $0.905$ (blue) and $0.91$ (red). Here, $\dot{m} = 0.01$, $\alpha_{\rm B} = 0.01$, $\mathcal{E}_{\rm in} = 5.8 \times 10^{-3}$ and $\lambda_{\rm in} = 2.34$, respectively.  See the text for details.}
	\label{spin}
\end{figure}

It is useful to study the role of black hole spin ($a_{\rm k}$) in generating the outflows from an accretion disc. For this, we compute mass outflow rate ($R_{\dot{ m}}$) by varying $a_{\rm k}$ values. Here, we choose the input parameters as $\mathcal{E}_{\rm in} = 5.8 \times 10^{-3}$, $\lambda_{\rm in} = 2.34$, $\dot{m} = 0.01$ and $\alpha_{\rm B} = 0.01$, respectively. The obtained results are illustrated in Fig. \ref{spin}, where dotted (green), dashed (blue) and solid (red) curves denote results corresponding to $a_{\rm k} = 0.90$, $0.905$ and $0.91$, respectively. Note that we consider marginal variation of $a_{\rm k}$ while comparing the $R_{\dot{ m}}$ values. This is done simply to ensure that the flow angular momentum $\lambda_{\rm in}$ renders self-consistent GIOS for the chosen range of $a_{\rm k}$ values. We find that for a fixed $\beta_{\rm in}$, $R_{\dot{ m}}$ increases with $a_{\rm k}$. In this analysis, we carry out the computation of $R_{\dot{ m}}$ keeping the input parameters fixed at the inner boundary ($i.e.$, at $x_{\rm in}$). Hence, when $a_{\rm k}$ is increased keeping $\lambda_{\rm in}$ fixed, the shock front recedes away from the horizon. This happens because of the spin-orbit coupling embedded in the effective potential describing the black hole spacetime, where marginally stable angular momentum anti-correlates with $a_{\rm k}$ \cite[]{Das-Chakrabarti2008}. As a result, the inflowing matter is intercepted by the enhanced effective area of PSC and increased $R_{\dot m}$ is resulted. 

\begin{figure}
	\includegraphics[width=\columnwidth]{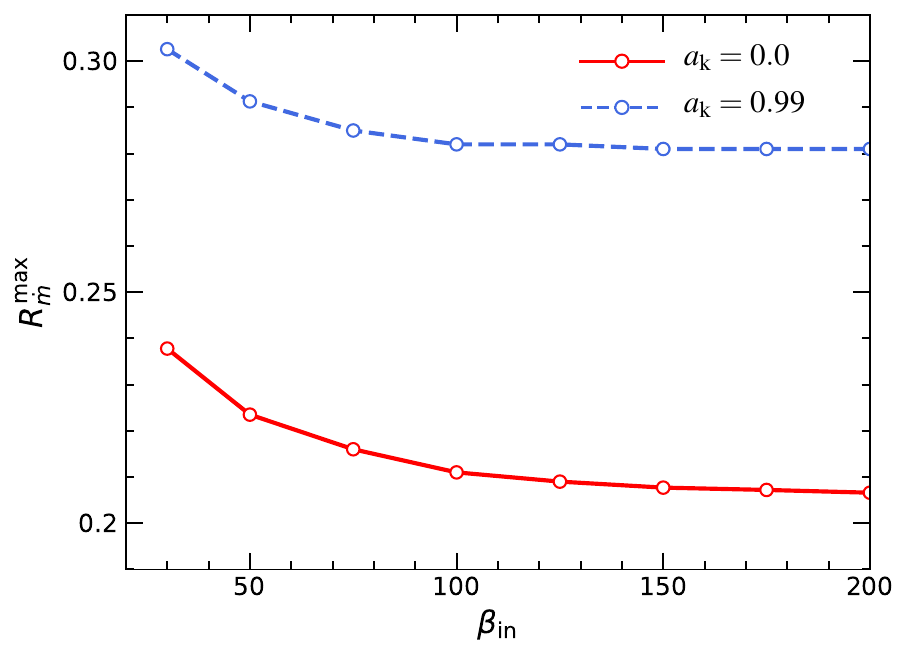}
	\caption{ Variation of maximum outflow rate ($R^{\rm max}_{\dot m}$) as function of $\beta_{\rm in}$ for $a_{\rm k} = 0$ (red) and $\rm a_k = 0.99$ (blue). Here, we choose $\alpha_{\rm B} = 0.005$ and $\dot{m} = 0.01$. See the text for details.}
	\label{rmax}
\end{figure}

Next, we put efforts to compute the limiting value of mass outflow rate, $i.e.$, the maximum value of $R^{\rm max}_{\dot m}$. While doing this, we keep the accretion rate and viscosity parameter fixed as ${\dot m}=0.01$ and $\alpha_{\rm B} = 0.005$, and freely vary the  $\mathcal{E}_{\rm in}$ and $\lambda_{\rm in}$ to calculate $R^{\rm max}_{\dot m}$ in terms of $\beta_{\rm in}$ and $a_{\rm k}$. The obtained results are depicted in Fig. \ref{rmax}, where the variation of $R^{\rm max}_{\dot m}$ is plotted with $\beta_{\rm in}$. Here, open circles joined with solid (in red) and dashed (in blue) lines are for $a_{\rm k} = 0$ and $0.99$, respectively. From the figure, it is evident that rapidly rotating black hole yields higher $R^{\rm max}_{\dot m}$ ($\sim 7\%$ higher) compared to the stationary black hole irrespective to $\beta_{\rm in}$ values. Noticeably, these results are in contrast with the results of gas dominated disc where $R^{\rm max}_{\dot m}$ exhibits marginal variation ($\sim 1\%$) with $a_{\rm k}$ \cite[]{aktar-etal-2015}. We further observe that $R^{\rm max}_{\dot m}$ reaches to $\sim 30\%$ ($24\%$) for $a_{\rm k} = 0.99$ ($0.0$). All these findings clearly suggest that magnetized accretion disc around highly spinning black hole is more likely to exhibit higher mass loss than the weakly rotating black hole. What is more is that for fixed $a_{\rm k}$, $R^{\rm max}_{\dot m}$ remains largely insensitive to the $\beta_{\rm in}$ except for low $\beta_{\rm in}$ domain.

\begin{figure}
	\includegraphics[width=\columnwidth]{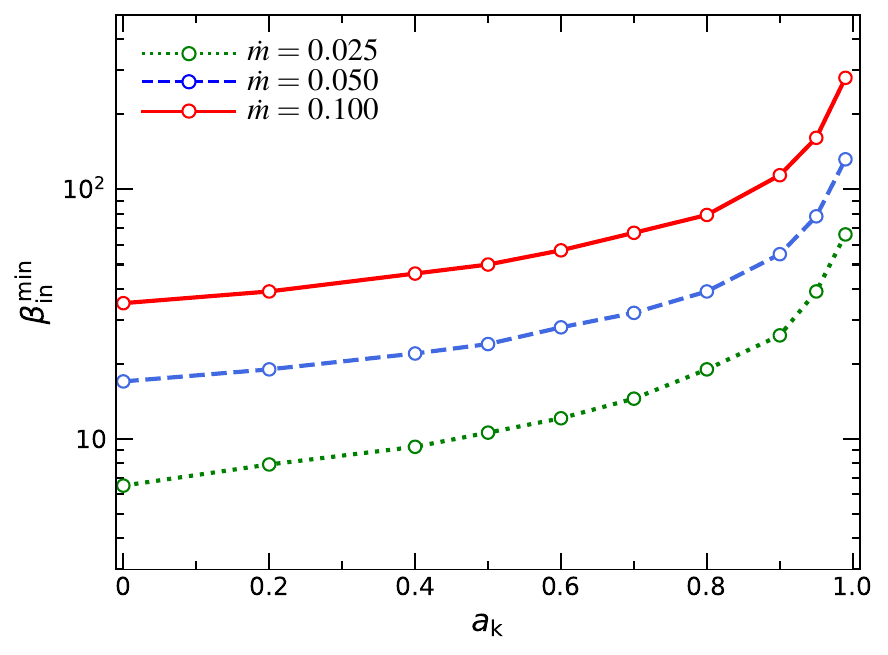}
	\caption{Variation of $\beta^{\rm min}_{\rm in}$ (minimum value of plasma-$\beta$ at $x_{\rm in}$) as function of $a_{\rm k}$ for different mass accretion rate ($\dot{m}$). Here, we choose $\alpha_{\rm B} = 0.005$. Open circles joined with dotted, dashed and solid lines denote results for ${\dot m}=0.025$ (green), $0.05$ (blue) and $0.1$ (red), respectively. See the text for details.
		}
	\label{akbetamin}
\end{figure}

In this work, the magnetic activity of the magnetized accretion disc is regulated using plasma-$\beta$ parameter. Therefore, it would be essential to analyse the maximally magnetized accretion disc that renders mass loss by quantifying the minimum value of plasma-$\beta$ ($i.e.$, $\beta^{\rm min}_{\rm in}$). Accordingly, we vary ${\cal E}_{\rm in}$ and $\lambda_{\rm in}$ freely and identify $\beta^{\rm min}_{\rm in}$ for a set of ($\alpha_{\rm B}$, $\dot m$, $a_{\rm k}$) yielding non-zero $R_{\dot m}$. For the purpose of representation, we choose $\alpha_{\rm B} = 0.005$ and find $\beta^{\rm min}_{\rm in}$ in terms of $a_{\rm k}$ and $\dot m$. The obtained results are depicted in Fig. \ref{akbetamin}, where open circles joined with dotted (green), dashed (blue) and solid (red) lines are for ${\dot m} = 0.025$, $0.05$ and $0.1$, respectively. Figure clearly indicates that mass loss from magnetized accretion disc around rotating black hole continues to happen for the spin range $0 \le a_{\rm k} < 1$. We find that for a given $\dot m$, $\beta^{\rm min}_{\rm in}$ increases with $a_{\rm k}$. More precisely, we notice that when black holes rotates slowly, outflows are generated from accretion disc even in presence of intense magnetic field, where magnetic pressure tends to become comparable with the gas pressure ($\beta^{\rm min}_{\rm in} \sim$ few $10$). On the contrary, the ejection of matter in the form of outflow is mostly possible from the gas pressure dominated disc (in presence of feeble magnetic fields) around rapidly rotating black holes. Moreover, we observe that for a fixed $a_{\rm k}$, $\beta^{\rm min}_{\rm in}$ increases with $\dot m$ indicating the fact that disc accreting at low rate can sustain more magnetic fields while deflecting matter as outflows.

So far, we have investigated the role of model parameters, namely ${\dot m}$, $\alpha_{\rm B}$, $\beta_{\rm in}$ and $\lambda_{\rm in}$ in regulating the mass outflow rate ($R_{\dot m}$) originated from the magnetized disc. Indeed, these outflows can explain the features of persistent radio emissions often observed from the galactic black hole sources (GBHs) in their low-hard state (LHS) and hard-intermediate state (HIMS). Keeping this in mind, in \S4, we make use of the coupled accretion-ejection formalism to elucidate the radio-jet power commonly observed from the BH systems.

\section{Astrophysical Implications}

\begin{table*}
	\caption{Comparison between observed and model predicted jet kinetic power. Columns 1-7 represent source name, source mass, distance, spin, observed X-ray flux, X-ray luminosity and mass accretion rate, respectively. Quantities in columns 8 and 9 indicate plasma-$\beta_{\rm in}^{\rm min}$ parameter and corresponding maximum outflow rate. Quantities in columns 10 and 11 denote model predicted maximum jet kinetic power and the observed jet kinetic power. In column 12, we provide references in order of $M_{\rm BH}$, $D$, $a_{\rm k}$, X-ray flux and radio flux, respectively, except the sources marked with $\dagger$ for which the spin is unknown.}
\centering{
	\renewcommand{\arraystretch}{1.3}
	\resizebox{2.15\columnwidth}{!}{%
		\begin{tabular}{@{}lccccccccccc}
			\hline
			Source & $ M_{\rm {BH}}$ & D& $ a_{\rm k}$ &$\rm F_{\rm X}$&$\rm L_{\rm X}$&$ \dot{m}$& $\beta_{\rm {in}}^{\rm min}$ & $ R_{ \dot{m}}^{\rm max}$ &  $ L_{\rm jet}^{\rm max}$ &  $L_{\rm jet}^{\rm obs}$& References \\
			
			& ($M_{\odot}$) &(kpc)&  &($10^{-10}$ erg $\rm s^{-1}$)&($10^{37}$ erg $\rm s^{-1}$)&($\dot{M}_{\rm Edd}$)  &   &&($10^{36}$erg $\rm s^{-1}$) &($10^{36}$erg $s^{-1}$)& \\ 
			\hline
			
			XTE J1550$-$564 & $9.1$&$4.4$ & $0.78$ &$57.8$ & $1.34$ & $0.039$  & $28$ & $0.258$ &$11.45$  & $6.31$& \cite{Orosz-etal-2011a},\cite{Orosz-etal-2011a},\cite{Miller-etal-2009},\cite{Gultekin-etal-2019},\cite{Gultekin-etal-2019}\\
			
			GRO J1655$-$40 & $6.3$ & $3.2$ & $0.98$ & $5.7$& $0.0070$ & $0.003$ &$4$ & $0.316$ &  $0.75$ & $1.61 $& \cite{Greene-etal-2001},\cite{Jonker-Nelemans-2004},\cite{Stuchlk-Kolo-2016},\cite{Gultekin-etal-2019},\cite{Gultekin-etal-2019}\\
			
			MAXI J1820$+$070 & $5.73-8.34$ & $2.96$ & $0.2$  &$276.5$& $2.90$ & $0.109$ & $43$ & $0.220$& $21.10$ &  $15.84 $ &\cite{Torres-etal-2020},\cite{Atri-etal-2020},\cite{Guan-etal-2021},\cite{Geethu-etal-2022},\cite{Trunshkin-etal-2018}\\

			GX 339$-$4 & $10.08$ & $8.4$ & $0.97$  &$2.0$& $0.17$ & $0.0045$ & $6$ & $0.283$ & $1.60$  & $3.00$&\cite{Sreehari-etal-2019b},\cite{Parker-etal-2016},\cite{Ludlam-etal-2015},[$^\boxtimes$],\cite{Corbel-etal-2021} \\ 

			Cyg X$-$1 & $14.8$ & $1.86$ & $0.99$ &$115.0$ & $0.48$ & $0.0085$ & $21$ & $0.295$ & $4.64$ & $3.81$ &\cite{Orosz-etal-2011b},\cite{Reid-etal-2011},\cite{Zhao-etal-2021,Kushwaha-etal-2021},\cite{Gultekin-etal-2019},\cite{Gultekin-etal-2019}\\
			
			IGR J17091$-$3624 & $10.6-12.3$&$11-17$ & $0.27$ &$11.4$& $2.67$&$0.062$& $25$ & $0.239$ & $21.22$  &$13.81 $ &\cite{Iyer-etal-2015},\cite{Rodriguez-etal-2011},\cite{Wang-etal-2018},\cite{Rodriguez-etal-2011},\cite{Rodriguez-etal-2011}\\ 
			
			XTE J1859$+$226 & $6.55$ & $6-11$ & $0.6$  & --- & $0.45$&$0.018$ &$8$ & $0.266$ &  $3.92$  & $2.30$ &\cite{Nandi-etal-2018},\cite{Hynes-etal-2002,Zurita-etal-2002},\cite{Steiner-etal-2013},\cite{Merloni-etal-2003},\cite{Merloni-etal-2003}\\
			
			MAXI J1348$-$630$^{\dagger}$ & $ 11 $ & $2.2$ & $0$  & --- & $3.60$ & $0.087$ &$31$& $0.221$ &  $26.46 $ & $6.60 $ & \cite{Lamer-etal-2021},\cite{Chauhan-etal-2021},\cite{Carotenuto-etal-2021},\cite{Carotenuto-etal-2021}\\
			
			& $ 11 $ & $2.2$ & $0.99$  & --- & $3.60$ & $0.087$ &$236$& $0.280$ &  $33.52 $ & $6.60 $ & \cite{Lamer-etal-2021},\cite{Chauhan-etal-2021},\cite{Carotenuto-etal-2021},\cite{Carotenuto-etal-2021}\\
						
			MAXI J1535$-$571 & $ 6.47 $ & $4.1$ & $0.99$  & --- & $12.4$ & $0.510$ & $1394$& $0.278$ & $ 114.76$ & $29.00 $&\cite{Sreehari-etal-2019},\cite{Chauhan-etal-2019},\cite{Miller-etal-2018},\cite{Russell-etal-2019a},\cite{Russell-etal-2019a}\\
			
			MAXI J0637$-$430$^{\dagger}$ & $ 8 $ & $10$ & $0$ &  --- & $2.25$ & $0.075$ & $26$ & $0.232$ & $17.41$ & $0.79 $&\cite{Baby-etal-2021},\cite{Tetaremko-etal-2021},\cite{Russell-etal-2019c},\cite{Russell-etal-2019c}\\
			
			& $ 8 $ & $10$ & $0.99$ &  --- & $2.25$ & $0.075$ & $204$ & $0.280$ & $21.02$ & $0.79 $& \cite{Baby-etal-2021},\cite{Tetaremko-etal-2021},\cite{Russell-etal-2019c},\cite{Russell-etal-2019c}\\
						
			V404 Cyg & $ 9.0 $ & $2.39$ & $0.92$  & --- & $0.82$ & $0.024$ &$29$& $0.286$ &  $ 7.73$ & $15.21 $&\cite{Khargharia-etal-2010},\cite{Miller-Jones-2009},\cite{Walton-etal-2017},\cite{Plotkin-etal-2019},\cite{Plotkin-etal-2019}\\
			
			H 1743$-$322 & $11.21$ & $8.5$ & $0.7$ & --- & $1.32$ & $0.031$ &$19$  & $0.267$ &  $11.61$  & $7.44$& \cite{Molla-etal-2017},\cite{Steiner-etal-2012},\cite{Steiner-etal-2012},\cite{Williams-etal-2020},\cite{Williams-etal-2020}\\ 
			
			\hline
			
		\end{tabular}	
	}	
}
\label{table}
\justify $^\boxtimes$Aneesha et al. (2024)(under review).\\ $^{\dagger}$Spin is not constrained, and hence, both non-rotating ($a_{\rm k}=0.0$) and rapidly rotating ($a_{\rm k}=0.99$) limits are considered.

\end{table*}

In this section, we attempt to explain the observed jet luminosity using our theoretical model formalism. Since the present work pertains to steady outflows, we focus on those sources where persistent jets are observed. Indeed, persistent jets are generally observed in the low-hard spectral states (LHS) of the galactic black holes (GBHs) \cite[]{Gallo-etal2003,Fender-etal-2004, Fender-etal-2009, Radhika-Nandi-2014,Radhika-etal-2016}. These steady jets are generally compact and yet to be separated from the central source \cite[]{Mirabel-Rodriguez1994,Corbel-etal2001,Fender-etal2001}. Keeping this in mind, we select $12$ BH-XRBs in LHS provided their simultaneously or quasi-simultaneously X-ray and radio observations are readily available. Moreover, the physical parameters of these sample sources, namely mass ($M_{\rm BH}$), distance ($D$) and spin ($a_{\rm k}$) are constrained (see Table \ref{table}). Utilizing all these, we estimate the jet kinetic power ($L_{\rm jet}$) adopting the following approach.

We calculate the mass accretion rate $\dot{M}_{\rm acc}$ (equivalently ${\dot M}_{-}$) of the black hole as  
\begin{equation}
\dot{M}_{\rm acc} = \frac{L_{\rm X}}{\eta_{\rm acc}c^2} \quad \rm g~s^{-1},
\label{acc}
\end{equation}
where $L_{\rm X}$ denotes the X-ray luminosity and $\eta_{\rm acc}$ refers the accretion efficiency factor. We obtain the X-ray luminosity from the literature for most of the sources, and for the remaining sources, we estimate source luminosity as $L_{\rm X} = 4 \pi D^2 F_{\rm X}$ knowing the X-ray flux ($F_X$, $1-10$ keV) from the observations, where $D$ being the source distance. We express the mass accretion rate in unit of Eddington rate ($\dot{ M}_{\rm Edd} = 1.39 \times 10^{17} M_{\rm BH}/M_\odot$ g s$^{-1}$)
\begin{align*}
	\dot{m} =\frac{{\dot M}_{\rm acc}}{{\dot M}_{\rm Edd}}&= 2.398 \times 10^{-17} \left(\frac{L_X}{c^2} \right) \left( \frac{M_{\rm BH}}{M_\odot}\right)^{-1} \\
	&= 3.01 \times 10^{-16} \Big( \frac{F_{\rm X} D^2}{c^2}\Big) \Big(\frac{M_{\rm BH}}{M_{\odot}}\Big)^{-1},
\end{align*}
where we consider $\eta_{\rm acc} = 0.3$ yielding the maximum radiative efficiency \cite[]{Throne-1974}.

Meanwhile, we notice that for a fixed $a_{\rm k}$, mass outflow rate ($R_{\dot m}$) increases as the magnetic activity inside the disk is increased (see Fig. \ref{rmax}). Motivating with this, we intend to compute the maximum outflow rate ($R^{\rm max}_{\dot m}$) (equation (\ref{rmdot})) from a maximally magnetized accretion disk while explaining the observed jet luminosity. In doing so, we employ the accretion-ejection model formalism and obtain $R^{\rm max}_{\dot m}$ corresponding to the minimum value of plasma-$\beta$ ($\beta^{\rm min}_{\rm in}$) at $x_{\rm in}$ by freely varying energy (${\cal E}_{\rm in}$) and angular momentum ($\lambda_{\rm in}$) of the flow. Here, we choose $\alpha_{\rm B} = 0.005$. Subsequently, using $\dot m$ and $a_{\rm k}$ for a given black hole, we compute the maximum jet kinetic power ($L^{\rm max}_{\rm jet}$) from the theoretical model as
\begin{equation}
	L^{\rm max}_{\rm jet} =  R^{\rm max}_{{\dot{ m}}} \, \dot{ M}_{\rm acc} \, c^2 \rm  ~erg~s^{-1}.
	\label{jet}
\end{equation}

Next, we compare $L^{\rm max}_{\rm jet}$ with observation. While doing so, 
we estimate the jet power from the radio luminosity using the empirical relation \cite[]{Blandford-Konigl1979,Heinz-Sunyaev2003,Heinz-Grimm-2005,Huang-etal-2014} given by,
\begin{equation}
	L^{\rm obs}_{\rm jet} = 4.79 \times 10^{15} L^{12/17}_{\rm R} \quad \rm erg\hspace{0.1cm} s^{-1},
	\label{L-obs}
\end{equation}
where $L_{\rm R}$ denotes radio luminosity computed using radio flux ($F_{\nu}$) measured at frequency $\nu$ as $L_{\rm R} = 4 \pi D^2 \nu F_{\nu}$. Here, radio data has been taken at a 
frequency $\nu \sim 5$ GHz. In this work, we obtain $L_{\rm R}$ from the literature and using equation (\ref{L-obs}), we calculate $L^{\rm obs}_{\rm jet}$ for a given black hole. In Table \ref{table}, we tabulate the details of the selected sources, where columns 1-12 denote source name, mass ($M_{\rm BH}$), distance ($D$), spin ($a_{\rm k}$), X-ray flux ($F_X$), X-ray luminosity ($L_X$), accretion rate ($\dot m$), $\beta^{\rm min}_{\rm in}$, $R^{\rm max}_{\dot m}$, $L^{\rm max}_{\rm jet}$, $L^{\rm obs}_{\rm jet}$ and relevant references, respectively. Note that for few sources, the spin parameters are not known constrained and hence, we estimate $L^{\rm max}_{\rm jet}$ considering the limiting values, such as $a_{\rm k} \rightarrow 0$ (weakly rotating) and $a_{\rm k} =0.99$ (rapidly rotating). For MAXI J1820$+$070 and IGR J17091$-$3624, we consider average mass calculated using their available mass range. Similarly, for IGR J17091$-$3624 and XTE J1859$+$226, the distance is not well constrained and hence, we use their average value. It is evident from Table \ref{table} that for most of the sources, namely XTE J1550$-$564, MAXI J1820$+$070, Cyg X$-$1, IGR J17091$-$3624, XTE J1859$-$226, MAXI J1348$-$630, MAXI J1535$-$571, MAXI J0637$-$430 and H 1743$-$322, the observed jet kinetic power ($L^{\rm obs}_{\rm jet}$) lies within the theoretical estimates of maximum jet kinetic power ($L^{\rm max}_{\rm jet}$). For the remaining sources, such as GRO J1655$-$40, GX 339$-$4 and V404 Cyg, $L^{\rm max}_{\rm jet}$ tends to agree with $L^{\rm obs}_{\rm jet}$ within the same order of estimates. With this, we argue/indicate that the present accretion-ejection model formalism seems to be potentially viable to explain the radio jet power of these selected sources. What is more is that the accretion disk around GRO J1655$-$40, GX 339$-$4 and XTE J1859$+$226 are appears to be strongly magnetized ($\beta^{\rm min}_{\rm in} < 10$), whereas MAXI J1535$-$571 seems to gas pressure dominated ($\beta^{\rm min}_{\rm in} > 100$). 

\section{SUMMARY and CONCLUSION}

In this work, we study mass loss in the form of the outflows from a magnetized, viscous, advective accretion disk around a rotating BH in presence of synchrotron cooling, for the first time to the best of our knowledge. While doing this, we consider the accretion disk to be threaded by the toroidal magnetic fields \cite[]{Oda-etal-2007,Sarkar-Das2016,Sarkar-etal2018} and also confined around the BH equatorial plane. Depending on the model parameters, such a disk may contain centrifugally supported shocks yielding a post-shock corona (PSC) surrounding the BH, where a part of the magnetized accreting matter is deflected to produce bipolar outflows. These outflows are emerged out from the disk along the rotational axis of BH guided by the funnel wall and centrifugal barrier \cite[]{Molteni-etal1996}. Further, in order to avoid the general relativistic complexity, we adopt a recently developed effective potential \cite[]{Dihingia-etal-2018} that satisfactorily mimics the spacetime geometry around the rotating BH. The main findings of this study are summarized below.

\begin{itemize}
	
	\item We compute the mass outflow rate ($R_{\dot m}$) from a magnetized accretion flow around the rotating BHs by solving the coupled accretion-ejection equations self-consistently. In order to examine the effect of magnetic fields on the matter ejection process, we introduce plasma-$\beta$ parameter defined as the ratio of gas pressure to magnetic pressure. We observe that magnetized accretion disk continues to eject matter in the form of outflow for wide ranges of model parameters, namely accretion rate ($\dot m$), viscosity ($\alpha_{\rm B}$), angular momentum of the flow ($\lambda$), spin of the black hole ($a_{\rm k}$) and magnetic fields (plasma-$\beta$).
	
	\item We notice that for a set of model parameters, the mass outflow $R_{\dot m}$ increases as the magnetic activity is increased inside the disk (see Figs. \ref{mdot}, \ref{lambda}, \ref{alpha}, \ref{spin}).
	
	\item We estimate the maximum mass outflow rate ($R^{\rm max}_{\dot m}$) from a magnetized disk and find that $R^{\rm max}_{\dot m}$ remains always higher for rapidly rotating black hole ($a_{\rm k} \rightarrow 0.99$) compared to the stationary black hole ($a_{\rm k} = 0.0$) irrespective to the plasma-$\beta$ parameter. Moreover, we observe that for magnetic pressure dominated disk, 	$R^{\rm max}_{\dot m}$ reaches up to $\sim 30\%$ for $a_{\rm k} \rightarrow 0.99$, whereas $R^{\rm max}_{\dot m}\sim 24\%$ for $a_{\rm k} \rightarrow 0$ (see Fig. \ref{rmax}).
	
	\item We analyse the maximally magnetized disk (parametrized with $\beta^{\rm min}_{\rm in}$)  around BH that renders outflow. We find that for slowly rotating BH ($a_{\rm k} \rightarrow 0$), accretion flow threaded with intense magnetic field ($\beta^{\rm min}_{\rm in} \sim {\rm few}~10$) admits mass loss, whereas the outflows are likely to launch from the vicinity of rapidly rotating BH ($a_{\rm k} \rightarrow 1$) for relatively large $\beta^{\rm min}_{\rm in}$ (see Fig. \ref{akbetamin}).
	
	\item We use the accretion-ejection formalism to explain the observed jet kinetic power ($L^{\rm obs}_{\rm jet}$) of several BH-XRBs in their low-hard spectral states. Employing our theoretical model formalism, we compute maximum jet kinetic power ($L^{\rm max}_{\rm jet}$) and find that $L^{\rm max}_{\rm jet}$ for the selected sample sources are in agreement with $L^{\rm obs}_{\rm jet}$ (see Table \ref{table}).

\end{itemize}

Finally, it is essential to mention the limitations of this work. We consider an effective potential to describe the spacetime geometry around a rotating black hole instead of using proper general relativistic treatment. We assume the accretion disk to be threaded by the toroidal magnetic fields neglecting the poloidal components, and also ignore magnetic fields in the outflows. We further use adiabatic index ($\gamma$) as a global constant, rather than calculating it self-consistently based on the temperature profile of the flow following relativistic equation of state. Moreover, we consider only synchrotron cooling process neglecting bremsstrahlung emission and Compton emission processes. Indeed, all these physical processes are relevant in the context of the accretion-ejection mechanism, and hence, we plan to take up these issues as future works and will be communicated elsewhere.

\section*{Data Availability}

The data underlying this paper will be available with reasonable request.

\section*{Acknowledgements}

Authors thank the Department of Physics, IIT Guwahati, India for providing the infrastructural support to carry out this work.


\section*{Appendix}

Using equation (\ref{mdot}) in equations (\ref{momenacc}), (\ref{angular}), (\ref{entropy}) and equation (\ref{magflux}) in equation (\ref{fluxvary}), we get

\begin{subequations}
	\begin{equation}
		\mathcal{E}_v \frac{dv}{dx} + \mathcal{E}_a \frac{da}{dx} + \mathcal{E}_{\lambda} \frac{d\lambda}{dx} + \mathcal{E}_{\beta} \frac{d\beta}{dx} + \mathcal{E}_0 = 0
		\label{EE}
	\end{equation}
	
	\begin{equation}
		l_v \frac{dv}{dx} + l_a \frac{da}{dx} + l_{\lambda} \frac{d\lambda}{dx} + l_{\beta}\frac{d \beta}{dx} + l_0= 0
		\label{ll}
	\end{equation}

	\begin{equation}
		R_v \frac{dv}{dx} + R_a \frac{da}{dx} + R_{\lambda} \frac{d\lambda}{dx} + R_{\beta}\frac{d \beta}{dx} + R_0= 0
		\label{RR}
	\end{equation}

	\begin{equation}
		b_v \frac{dv}{dx} + b_a \frac{da}{dx} + b_{\lambda} \frac{d\lambda}{dx} + b_{\beta}\frac{d \beta}{dx} + b_0= 0
		\label{bb}
	\end{equation}

\end{subequations}

The coefficients of equations (\ref{EE} - \ref{bb}) are expressed in the form of
\\

$\mathcal{E}_{v} = \Big(\frac{\gamma v^2 - a^2}{\gamma v}\Big)$, $\mathcal{E}_{a} = a/\gamma$, $\mathcal{E}_{\lambda} = \Big(\frac{a^2}{2 \gamma \mathcal{F}}\frac{\partial \mathcal{F}}{\partial \lambda}\vert_{x}\Big)$, $\mathcal{E}_{\beta} = 0$ and $\mathcal{E}_{0} = \frac{a^2}{2 \gamma \mathcal{F}}\frac{\partial \mathcal{F}}{\partial x}\vert_{\lambda} - \frac{3 a^2}{2 \gamma x} - \frac{a^2 {\Delta}^{\prime}}{2 \gamma \Delta} + \frac{d \Psi^{\rm eff}}{dx} + \frac{2 a^2}{\gamma (1 + \beta) x}; \Delta^{\prime} = \frac{d\Delta}{dx} = 2 (x-1)$
\\

 $l_v = \alpha_{\rm B} x \Big(1 - \frac{g a^2}{\gamma v^2}\Big)$, $g = \frac{I_{n+1}}{I_n}$, $l_a = \frac{2 \alpha_{\rm B} x g a}{\gamma v}$, $l_{\lambda} = -1$, $l_{\beta} = 0$ and $l_0 = \frac{\alpha_{\rm B}}{\gamma v}(g a^2 + \gamma v^2)(2 - \frac{x \Delta^\prime}{2 \Delta})$ 
\\

 $R_v= \frac{a^2}{\gamma}\frac{\beta}{1+ \beta}$, $R_a=\frac{\gamma+1}{\gamma-1}\frac{a^2 v}{\gamma}\frac{\beta}{1+ \beta}$, $R_{\lambda} = -\Big(\frac{a^2 v}{\gamma}\frac{\beta}{1+ \beta}\frac{1}{2 \mathcal{F}}\frac{\partial \mathcal{F}}{\partial \lambda}\vert_{x}  + \frac{2 \alpha_{B} I_n}{\gamma} (g a^2 + \gamma v^2) x \frac{\partial \Omega}{\partial \lambda}\vert_{x}\Big)$, $R_\beta = \frac{a^2 v}{\gamma (\gamma-1)(1 + \beta)^2}$ and $R_0 = \frac{a^2 v}{\gamma}\frac{\beta}{1+ \beta}\Big(\frac{\Delta^\prime}{2 \Delta} + \frac{3}{2 x} -\frac{1}{2 \mathcal{F}}\frac{\partial \mathcal{F}}{\partial x}\vert_{\lambda}\Big) - \frac{s a^5}{v}\sqrt{\frac{\mathcal{F}}{x^3 \Delta}}\frac{\beta ^2}{(1 + \beta)^3} - \frac{2 \alpha_{B} I_{n}}{\gamma}(g a^2 + \gamma v^2) x \frac{\partial \Omega}{\partial x}\vert_{\lambda}$
  \\

$b_v = 1/v$, $b_a = 3/a$, $b_{\lambda} = \frac{-1}{2 \mathcal{F}}\frac{\partial \mathcal{F}}{\partial \lambda}\vert_{x}$, $b_{\beta} = -1/(1+ \beta)$ and $b_0 = \frac{2 \zeta}{x} - \frac{\Delta^{\prime}}{2 \Delta} + \frac{3}{2x} -\frac{1}{2 \mathcal{F}}\frac{\partial \mathcal{F}}{\partial x}\vert_{\lambda}$\\

The coefficients mentioned in equations (\ref{dadx}), (\ref{dldx}) and (\ref{dbetadx})
 are as follows,
 $a_{11} = -\frac{\mathcal{E}_{\lambda} l_0 + \mathcal{E}_0}{\mathcal{E}_{a} + \mathcal{E}_{\lambda} l_{a}}$, $a_{12} = -\frac{\mathcal{E}_{\lambda} l_v + \mathcal{E}_v}{\mathcal{E}_{a} + \mathcal{E}_{\lambda} l_{a}}$\\
$\lambda_{11} = a_{11} l_a +l_0$, $\lambda_{12} = a_{12} l_a + l_v$\\
$\beta_{11} = - \frac{b_0 + b_{\lambda} \lambda_{11} + b_a a_{11}}{b_{\beta}}$, 
$\beta_{12} = -\frac{b_v + b_{\lambda} \lambda_{12}+ b_a a_{12} }{b_{\beta}}$\\

By utilizing the aforementioned coefficients, the numerator and denominator of equation (\ref{dvdx}) can be expressed as

${\cal N}(x, v, a, \lambda, \beta)  = - (R_0 + R_a a_{11} + R_{\lambda} \lambda_{11} + R_{\beta} \beta_{11})$

${\cal D}(x, v, a, \lambda, \beta)  =  (R_v + R_a a_{12} + R_{\lambda} \lambda_{12} + R_{\beta} \beta_{12})$

\end{document}